\documentclass{article}
\usepackage{amsmath}
\usepackage[bitstream-charter]{mathdesign}
\usepackage{graphicx}
\usepackage{arxiv}
\usepackage{bm}
\usepackage{biblatex}
\bibliography{paxos.bib}

\numberwithin{equation}{section}

\begin{document}

\title{Understanding Paxos and other distributed consensus algorithms}

\author{Victor Yodaiken
\thanks{\copyright Victor Yodaiken, 2018-2021. }
}

\date{Feb 3 2018}
\keywords{consensus, Paxos}

\begin{abstract}
Paxos \cite{paxos-made-simple} is a widely used and notoriously  hard to understand\cite{Ongaro}  method for solving one type of distributed consensus problem. This note provides a quick explanation of Paxos, a novel proof of correctness that is intended to provide
insight into why the algorithm is as simple as the author has claimed, an explanation of  why it does and why it doesn't work, and has a brief discussion of alternatives.
\end{abstract}

\maketitle
\section{Paxos}
\begin{verse}
There was an old lady who swallowed a bird;\\ 
How absurd to swallow a bird! --- Bonne/Mills.
\end{verse}
Despite being not "live",  the Paxos algorithm \cite{paxos-made-simple, lamport-part-time, lamportpatent1,lamportpatent2}is a brilliantly intricate distributed consensus algorithm
and incorporates an interesting insight about the importance of correctly posing the problem.
Additionally, the algorithm, or related algorithms, are apparently widely used in practice according to papers like \textit{e.g.} \cite{paxoslive,bolosky,prisco} and indicators such as the  312 hits turned up when searching for "Paxos" and "consensus" in the USPTO database of issued patents. The basic algorithm is notoriously hard to understand\cite{Ongaro}, but is shown here to be really
as simple as the author has claimed. 

\subsection{Consensus}
Distributed consensus involves getting a collection of
agents (devices or processes or equivalent )  that communicate only by exchanging discrete data packets (messages) to  agree on a value. Consensus can involve true unanimous consensus or a 
majority or some other "quorum" depending on the specific requirements.
For example, a collection of database "replicas" might need to 
agree on when a transaction or series of transactions can be committed so that a distributed database stays consistent and fault tolerant.
If networks provide
reliable atomic broadcast the problem is
easy (see, for example the Auragen fault tolerant message systems \cite{borg}), but most work in this area assumes packets
can be delayed or lost or reordered  but that no malicious, spurious, or corrupted packets are delivered. 

\subsection{The naive method\label{sec:naive}}
If there is a single "proposer" agent (to propose a value), distributed consensus can be simple. The proposer sends a proposal message to 
all the "acceptors" and keeps resending until it has received acknowledgments from a quorum.
In most cases, there are not very many acceptors (often just 2 or 3), agents rarely fail, and  message delivery is fairly reliable and fast, so this process can be efficient.
If messages get through with 
high probability after a number of resends, the algorithm is highly probable to complete (to be "live"). The proposer can simply provide a new value at each round, possibly with a sequence number, to 
produce a sequence of agreed on values. 
There is a smart optimization method from Chang/Maxemchuck\cite{Chang} that is not widely used for some reason but can radically reduce message counts.

If the proposer fails a new unique proposer must be elected, but the 
election must ensure that the agents can come to consensus about  which proposer has won.  We have to solve the distributed consensus problem a second time to make
the first solution work in the presence of failures\footnote{Effectively, we have to swallow the bird to catch the spider.}. 
We'll return to this topic in section \ref{sec:time}. 

\section{Paxos 1 and 2: fly, spider, bird, otter, kangaroo, jellyfish, etcetera.}
The paper "Paxos made simple"\cite{paxos-made-simple} starts by providing a method with multiple competing proposers to build in resilience against proposer failures\footnote{I cannot decipher the original Paxos paper\cite{lamport-part-time}}.
The competition is based on the clever observation that we don't really care which proposer wins or whether multiple proposers win, we
only care that no more than a single consensus value wins. Proposers attempt two stages: in stage one they send acceptors a "pre-proposal" that only contains a proposer sequence number. In stage two proposers send a full proposal 
pair consisting of the same proposer sequence number and a \emph{proposed value}. No two proposers ever share a sequence number and a proposer cannot use the same sequence number
with two different values. A proposer starts stage 2 only when or if it has received acks for its pre-proposal from more than 
half of the acceptors and it wins only if it gets acks for its full proposal from more than half the acceptors.  An acceptor's acknowledgments are in strict order, each is either before or after any other acknowledgment by the same acceptor and 
an acceptor can only acknowledge proposals and pre-proposals if it has not previously acknowledged a greater sequence number either in 
a pre-proposal or a full proposal \footnote{This is a critical point and it is ambiguous in the text of "Paxos made simple". At one point the requirement as written is implied, at another point it is implied that proposers can only
send proposals to acceptors that have acknowledged the pre-proposal -- which would have the same effect.}.
Acceptors must attach the highest numbered full proposal the acceptor has already acknowledged to any pre-proposal acknowledgment. 
When it goes to the second stage, the proposer must
inherit (adopt) the highest numbered value from any proposals (if any) attached to the acks it was sent by acceptors. 
If no full proposal is sent back, the proposer  can chose an arbitrary value. All of this guarantees that 
the lowest numbered winning proposal determines the value for any other winning proposals - in fact, it determines the value
for any higher numbered proposals that are acknowledged even one time. We can prove this by induction.  

Suppose \((n_1,v_1)\) is the lowest numbered winning
proposal  and  list in order of sequence numbers all the greater or equal numbered proposals that have been acknowledged at least once \((n_1,v_1),\dots (n_k,v_k)\).
This list includes but is not limited to all proposals that have won - we just need one acknowledgment to put the proposal on the list after the first pair. It doesn't even matter whether the proposer received the acknowledgment. 
If \(j=1\), \(v_j = v_1\) by construction. Suppose \(j>1\) and for all \(1\leq i < j\), the list element \((n_i,v_i)\) has \(v_i = v_1\)  (induction hypothesis) and then consider the next list element \((n_j,v_j)\). 
For \((n_j,v_j)\) to be on the list it must have been acknowledged by at least one acceptor and the pre-proposal \(n_j\) had to have been acknowledged by
more than 1/2 of the acceptors. Since \((n_1,v_1)\) was acknowledged by more than 1/2 of the acceptors, at least one acceptor \(a\) has
to have acknowledged both \(n_j\) and \((n_1,v_1)\).
Since \(n_j > n_1\), it must be that \(a\) acknowledged \((v_1,n_1)\) before it acknowledged \(n_j\) so \(a\) sent 
back either \((n_1,v_1)\) or a higher numbered proposal to the proposer of \(n_j\). It follows that the proposer of \(n_j\) 
had to inherit some proposal \((n_i,v_i)\) with \(n_i \geq n_1\).
 It must be that \((n_i,v_i)\) is on the list since it had to have been acknowledged to be
sent back with the acknowledgment of \(n_j\). If acceptor \(a'\) sent \((n_i,v_i)\) back to the proposer of \(n_j\), \(a'\)
must have acknowledged 
\((n_i,v_i)\) before it acknowledged \(n_j\) so \(n_i < n_j\).  
But if \((n_i,v_i)\) is on the list with \(n_1\leq n_i< n_j\) then  \(v_i = v_1\) by the induction hypothesis. QED


Although Paxos is safe it is not live: the proposers can block 
each other and prevent any proposal value at all from winning. The proof of safety above gives clues about how easy it is for one proposal to block another.
Suppose proposer 1 with sequence number 1 succeeds in stage one and then proposer 2 with sequence number 2 succeeds in stage one and either fails or stalls indefinitely so
that proposer 1 cannot get any acks for its proposal. The proposed solution in "Paxos made simple" is  to allow proposers that cannot advance to increase their sequence number (the new number still needs to be unique), abandon the old 
proposal value, if any, and start over again. Basically, this doesn't change much - it is as if new proposers entered the competition and the one with the abandoned number stalled forever.
Unfortunately, it turns out that it is still possible for proposers to keep blocking and restarting so that the algorithm spins forever, never reaching consensus:
\begin{quote}
    It’s easy to construct a scenario in which two proposers each keep issuing
a sequence of proposals with increasing numbers, none of which are ever chosen. Proposer p completes phase 1 for a proposal number n 1 . Another
proposer q then completes phase 1 for a proposal number n 2 > n 1 . Proposer
p’s phase 2 accept requests for a proposal numbered n 1 are ignored because
the acceptors have all promised not to accept any new proposal numbered
less than n 2 . So, proposer p then begins and completes phase 1 for a new
proposal number n 3 > n 2 , causing the second phase 2 accept requests of
proposer q to be ignored. And so on.
\cite{paxos-made-simple}\end{quote}
Or see the simulation in \cite{yodaikensim}.

The explication in
\cite{paxos-made-simple} then shifts  to a  very different method, confusingly 
also called "Paxos"   where there is exactly one  
 proposer -- a "distinguished" proposer.
\begin{quote}{ In normal operation, a single server is elected to be the leader, which acts as the distinguished proposer (the only one that tries to issue proposals) in all instances of the consensus algorithm}\end{quote}
 How this proposer is to be chosen is not
 completely specified.
 The distinguished proposer can carry out the two stage process  and is assured to win since there will be no other competitor (or it could simply use the procedure of section \ref{sec:naive}). Just to make this point as clearly as possible, the two stage competition doesn't accomplish any purpose if there is a single proposer, so reverting to a single proposer algorithm and letting some other algorithm solve the distributed consensus problem for choosing that 
single proposer means that the second version of Paxos is just an inefficient version of the simple algorithm in section \ref{sec:naive}.

\section{Time \label{sec:time}}
In the passage where multi-proposer version of Paxos is abandoned for the single proposer version, "Paxos made simple" has the following note:
\begin{quote}
    The famous result of Fischer, Lynch, and Patterson \cite{fischer} implies that a reliable algorithm for electing a proposer must use
either randomness or real time— for example, by using timeouts.
\end{quote}
This is striking because the original premise of the paper is that there are no timeouts:  "Agents operate at arbitrary speed".  Completely "asynchronous" models of networks were
popular in academic computer science in the era in which "Paxos made simple" was written for reasons that are not clear, at least to me.
The "theorem" of  \cite{fischer} is the tautological result that if it is impossible to  distinguish between a failed agent and a slow to respond agent (e.g. if "Agents operate at arbitrary speed"  )
then it is impossible to distinguish between a failed agent and a slow to respond agent, which the authors sensibly conclude means that no totally clock free distributed consensus algorithm can be made to 
work in the presence of dropped messages and agents that can fail. Consideration of this observation leads to the conclusion that the first version of Paxos could be made to be live
in the presence of agent failures if proposers had to wait for a timeout before restarting with a new number. 
If a stalled proposer had to wait \(t\) time units before restarting with a new number and \(t\) was enough time so that it was highly probable a competing proposer could complete the protocol, then the algorithm would be live  (probably).  

The simple consensus algorithm of \ref{sec:naive} could be improved
to allow recovery from failure of the single proposer using clocks as proposed in numerous pre-Paxos systems such as those of Liskov\cite{liskov,oki-liskov} (and clock technology has advanced a lot in 30 years \cite{yodaiken-wsts-2018}). The proposer could be required to send a "heartbeat" message every so often to remind the acceptors that it was still alive. The
heartbeat could be combined with e.g. messages proposing new consensus values for efficiency. Acceptors could timeout if they had not
heard from a proposer for too long and then wait some randomized time before proposing themselves as the new proposer. Acceptors could
accept the new proposer with a time limited lease which extends when the candidate confirms it won by sending a heart beat. See 
\cite{Chang} for a well worked out system including state recovery.

In practice, distributed algorithms, including those named Paxos (see \cite{paxoslive, bolosky} for example) rely on timeouts and leases and other time related mechanisms. 
Raft \cite{Ongaro} relies on timeouts as well.
\printbibliography

\end{document}